\documentclass[preprint,proceedings]{rmaa}

\newcommand{\D}{\discretionary{}{}{}}

\SetYear{2002}
\SetConfTitle{Galaxy Evolution: Theory and Observations}

\title{Globular Clusters and Galaxy Formation
}

\author{
  Duncan A. Forbes\altaffilmark{1}}

\altaffiltext{1}{Centre for Astrophysics \& Supercomputing,
                 Swinburne University, Mail\# 31, PO Box 218,
                 Hawthorn, VIC 3122, Australia
                 (dforbes\D{}@astro.\D{}swin.\D{}edu.\D{}au).}

\suppressfulladdresses

\listofauthors{D. A. Forbes}
\indexauthor{Forbes, D. A.}

\addkeyword{galaxies: evolution}
\addkeyword{galaxies: formation}
\addkeyword{globular clusters: general}

\begin{document}
\maketitle 

\boldabstract{Globular clusters provide a unique probe of galaxy
formation and evolution. Here I briefly 
summarize the known observational properties
of globular cluster systems. One re-occurring theme is that the globular
cluster systems of spirals and ellipticals are remarkably
similar. Photometry, and the limited spectra available, 
are consistent with
metal-poor clusters forming before the main spheroid component is
established and the metal-rich ones forming at the same time as
the spheroid in a burst of star formation. These observations are
compared to a model for globular cluster formation in a
$\Lambda$CDM hierarchical universe. One model result reported
here is that
S$_N$ is determined at early times and little affected by late
epoch mergers. 
}

Globular clusters (GCs) trace the star formation and chemical
enrichment history of their host galaxy. As such they provide a
unique probe of galaxy evolution. They are also simple stellar
systems and so may be easier to understand than the galaxies
themselves which can possess a complex mix of stellar
populations. 

\begin{figure}[!t]
  \includegraphics[width=\columnwidth]{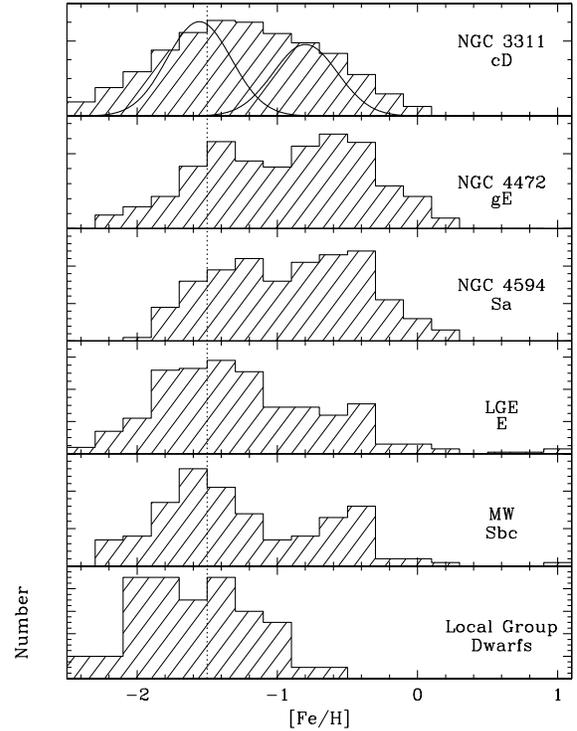}
  \caption{Globular Cluster metallicity distributions for a range
of Hubble types. All galaxies reveal a population of metal-poor
clusters ([Fe/H] $\sim$ --1.5). 
Local Group dwarfs with little or no bulge component
have very few metal-rich clusters, whereas all large galaxies
with a bulge/spheroid possess metal-rich clusters ([Fe/H] $\sim$ 
--0.5). 
}
  \label{Figure1}
\end{figure}

Good quality photometric data on GCs is now available for $\sim$ 50
E/S0s and a small number of spirals. 
In Fig.~\ref{Figure1} we show the GC metallicity distributions
for some galaxies with a range of Hubble types. 
Observations suggest that most 
moderate size galaxies have a bimodal 
GC metallicity distribution and that, to
first order, they appear with the same mean values, at [Fe/H]
$\sim$ --1.5, --0.5. This indicates the presence of two distinct 
GC subpopulations. 
Thus a simple one-stage collapse scenario
for elliptical galaxies can be ruled out (a two-phase
collapse remains a possibility; see Forbes, Brodie \& Grillmair
1997). Studies of the {\it inner metal-rich} GCs in the Milky Way
(eg Minniti 1995) and other spirals (Forbes, Brodie \& Larsen
2001) suggest that such GCs are associated with the bulge
component of spirals and not the disk. In this context, it is not
surprising that dwarf galaxies, with little or no bulge,
generally have few metal-rich GCs (see Fig.~\ref{Figure1}). 

Although all large galaxies have a metal-rich subpopulation,
the mean metallicity of this subpopulation 
varies with galaxy mass (Forbes \& Forte 2001). 
In other words the metal-rich GCs reveal a metallicity-mass 
relation. 
Interestingly, the metal-poor subpopulation appears to have a 
constant metallicity at
[Fe/H] $\sim$ --1.5. This suggests that the metal-poor GCs are
pre-galactic while the metal-rich ones know about the galaxy
potential they form in. These trends are true for ellipticals, S0s and 
the bulges of spirals. 

The GC systems of spirals and ellipticals share other common
properties. 
The number of metal-rich GCs per unit {\it bulge/spheroid} light
is roughly the same (with S$_N$ (bulge) $\sim$ 1) for spirals
and field ellipticals. 
Recently Larsen et al. (2001) showed that the mean 
sizes of individual GCs vary with metallicity for a range of
Hubble types. 
Furthermore the GC luminosity function peak and width are
remarkably constant between spirals and ellipticals. 
Thus the GC systems of spirals and ellipticals are more alike
than they are different, suggesting similar formation processes. 

Many presentations at this conference have focused on galaxy
formation in a $\Lambda$CDM hierarchical universe. Observations
of GCs provide another, alternative means in which test the
predictions of this model. One of the most basic predictions of
this picture is that the bulges of spirals and ellipticals 
formed in a similar way. Our observations of the similarities
in the GC systems of bulges and ellipticals would support this
view. 

Recently we simulated the formation of
GCs around ellipticals 
using the GALFORM semi-analytic code (eg Cole et al. 2000) in a
$\Lambda$CDM universe.  
Details of this work can be found in Beasley et al. (2002) and is the
subject of a poster at this conference by Beasley. Briefly, GCs
are formed in two modes of star formation. In the first, or
`quiescent' mode, metal-poor GCs form in proto-galactic
clouds. These gaseous clouds collapse/merge, giving rise to a
burst of star formation. During this `burst' mode, the vast bulk of
the galaxy stars form along with the metal-rich GCs. The final GC
system depends on the local galaxy environment and its mass
(luminosity). For example, low luminosity field ellipticals tend
to have a more extended star formation history which is more
bursty in nature. We would expect such galaxies to reveal
metal-rich GCs that are 3-5 Gyrs younger in the mean than the
metal-poor ones which are $\sim$12 Gyrs old in all ellipticals. 

\begin{figure}[!t]
  \includegraphics[width=\columnwidth]{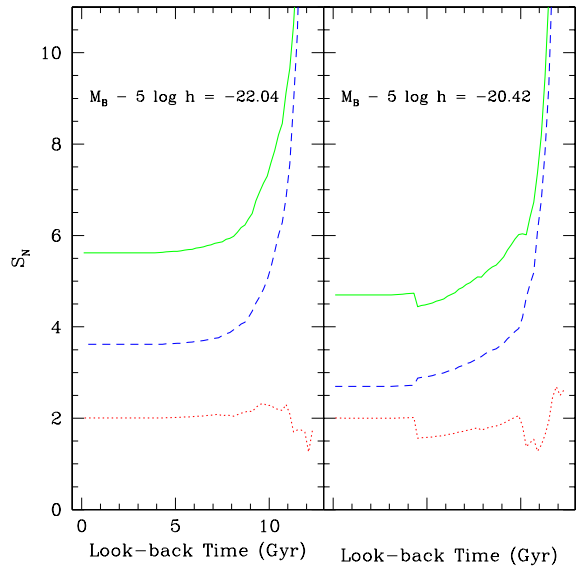}
  \caption{Evolution of globular cluster specific frequency S$_N$ with
look back time for two different galaxy realizations from
GALFORM. The dotted line shows the evolution of S$_N$ for the red
clusters, the dashed line for blue clusters, and the solid line
for the total cluster system. The lower luminosity galaxy undergoes a
merger at $\sim$5 Gyrs ago, which causes only a small increase in
the overall S$_N$ of the galaxy. 
}
  \label{Figure2}
\end{figure}

Another interesting aspect of our modeling is the evolution of GC
specific frequency S$_N$ (number of GCs per unit galaxy starlight).
In Fig.~\ref{Figure2} we show the evolution of S$_N$ for two
galaxies of different luminosities. The plots show that S$_N$ is
largely determined at early epochs and shows little or no change
in the last $\sim$8 Gyrs. In the case of the lower luminosity
galaxy, it undergoes a major merger $\sim$ 5 Gyrs ago. However
even though some new GCs are formed in that merger event, the
effect on S$_N$ is slight. 

In conclusion, the comparison of model predictions from $\Lambda$CDM and
observations of globular clusters 
are providing new insights into the formation
mechanism and evolution of galaxies. 

\end{document}